\documentclass[letterpaper,twocolumn,superscriptaddress,floatfix]{revtex4}
\usepackage{amsmath}
\usepackage{amsfonts}
\usepackage{amssymb}
\usepackage{graphicx}
\usepackage{bbm}
\usepackage{natbib}
\usepackage[dvipdfm]{hyperref}

\begin{document}

\title{Universal decay of quantumness for photonic qubits carrying orbital angular momentum through atmospheric turbulence}

\author{Mei-Song Wei}
\affiliation{Jiangsu Provincial Research Center of Light Industrial Optoelectronic Engineering and Technology, School of Science, Jiangnan University, Wuxi 214122,
China}
\author{Jicheng Wang}
\affiliation{Jiangsu Provincial Research Center of Light Industrial Optoelectronic Engineering and Technology, School of Science, Jiangnan University, Wuxi 214122,
China}
\author{Yixin Zhang}
\affiliation{Jiangsu Provincial Research Center of Light Industrial Optoelectronic Engineering and Technology, School of Science, Jiangnan University, Wuxi 214122,
China}
\author{Zheng-Da Hu}
\email{huyuanda1112@jiangnan.edu.cn}
\affiliation{Jiangsu Provincial Research Center of Light Industrial Optoelectronic Engineering and Technology, School of Science, Jiangnan University, Wuxi 214122,
China}
\date{\today}

\begin{abstract}
We investigate the decay properties of the quantumness including quantum entanglement, quantum discord and quantum coherence for two photonic qubits, which are partially entangled in their orbital angular momenta, through Kolmogorov turbulent atmosphere.
It is found that the decay of quantum coherence and quantum discord may be qualitatively different from that of quantum entanglement when the initial state of two photons is not maximally entangled.
We also derive two universal decay laws for quantum coherence and quantum discord, respectively, and show that the decay of quantum coherence is more robust than nonclassical correlations.
\end{abstract}
\maketitle

\section{Introduction}

The fact that photons can carry orbital angular momentum (OAM) to encode quantum states makes them very useful for quantum information science (QIS)~\cite{leach2002measuring,vaziri2003concentration,molina2007twisted,
nagali2009optimal,pors2011Highdimensional,fickler2012quantum}.
However, the decay of quantumness will be unavoidable when the encoded photons with OAM transmit through turbulent atmosphere.
Photons' wave front will be distorted due to refractive index fluctuations of the turbulent atmosphere,
which may lead to random phase aberrations on a propagating optical beam~\cite{paterson2005atmospheric}.
A large amount of efforts, both theoretical~\cite{gopaul2007effect,roux2011infinitesimal,sheng2012effects,brunner2013robust,gonzalez2013protecting} and experimental~\cite{pors2011transport,malik2012influence,da2013cancellation,rodenburg2014simulating},
have been devoted to exploring the impacts of atmospheric turbulence on the propagation of photons carrying OAM and protecting OAM photons from decoherence in turbulent atmosphere.

In QIS, it is typically critical to understand the behaviors of nonclassically correlated photons traveling in the turbulent atmosphere
since the quantumness contained in the encoded states are usually fragile and can be easily destroyed.
Quantum entanglement, a fundamental quantum resource in QIS, is a typical kind of quantumness~\cite{horodecki2009quantum}.
Recently, the decay of entanglement for photonic OAM qubit states in turbulent atmosphere has been reported~\cite{hamadou2013orbital,smith2006twophoton,
roux2015entanglement} via Wootters' concurrence~\cite{wootters1998entanglement}.
However, entanglement may not be the unique resource (or quantumness) which can be utilized in QIS.
There exist other resources such as quantum discord~\cite{Henderson2001,Ollivier2001,modi2012classical} and quantum coherence~\cite{baumgratz2014quantifying,girolami2014observable,Streltsov2017}  attracting much attention.
Recently, a measure of nonclassical correlation, termed as local quantum uncertainty (LQU)~\cite{girolami2013characterizing},
has been proposed as a genuine measure of quantum discord and is exactly computable for any bipartite quantum state.
The LQU, defined by the Wigner-Yanase skew information~\cite{wigner1963information}, is a quantification of the minimal quantum uncertainty achievable on a single local measurement, which is thus closely related to quantum metrology~\cite{girolami2013characterizing}.
On the other hand, a measure of coherence, defined by the relative entropy, has also been reported~\cite{baumgratz2014quantifying}. The relative entropy of coherence is shown to be a proper coherence quantifier, which not only fulfills all the conditions of coherence measure~\cite{Streltsov2017} but also is exactly computable for bipartite quantum state. It may be interesting to analytically explore the effect of turbulent atmosphere on these resources other than entanglement.

In this paper, we investigate the effects of atmospheric turbulence on the quantumness including quantum entanglement, quantum discord and quantum coherence for two photonic qubits of partially entangled in their OAM, via concurrence, LQU and relative entropy of coherence, respectively.
It is shown that the decay of quantum coherence and quantum discord may be qualitatively different from that of quantum entanglement when the initial state of two photons is not maximally entangled.
It is also found that the decay of quantum discord obeys an universal exponential law similar to that of entanglement already reported in Ref.~\cite{leonhard2015universal} but with asymptotic vanishing.
By contrast, the universal decay of quantum coherence is merely polynomial, indicating that quantum coherence is more robust against atmospheric turbulence.

This paper is organized as follows.
In Sec.~\ref{sec:sec2}, we derive the photonic OAM state influenced by the turbulent atmosphere and discuss the evolution of the entanglement of the initial extended Werner-like state.
In Sec.~\ref{sec:sec3}, the evolutions of the LQU and relative entropy of coherence as well as their decay laws are discussed.
Conclusions are presented in Sec.~\ref{sec:sec4}.

\section{Partially entangled OAM state through atmospheric turbulence}\label{sec:sec2}

\begin{figure}[tbh!]
\centering
\includegraphics[angle=0,width=8cm]{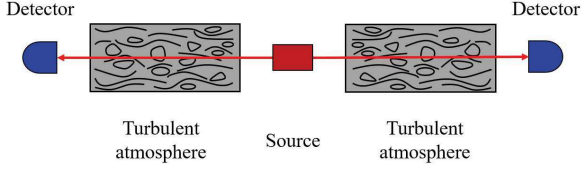}
\caption{A source produces pairs of OAM-entangled photons propagating through turbulent atmosphere and received by two detectors.}
\label{fig:fig1}
\end{figure}
In this work, we use two Laguerre-Gaussian (LG) beams~\cite{allen1992orbital} to generate a twin-photon state~\cite{fickler2012quantum}.
As is shown in Fig.~\ref{fig:fig1}, two correlated LG beams, generated by the source,
propagate through the turbulent atmosphere horizontally and then are received by the two detectors.
The beams carry pairs of photons that may not be maximally entangled in their OAM due to impurity and imperfection,
and the non-maximal entanglement is encoded by LG modes with the opposite azimuthal quantum number $l$~\cite{fickler2012quantum}.

We assume that the input LG modes have a beam waist $\omega_{0}$, a radial quantum number $p_{0}=0$, and azimuthal quantum numbers $l_{0}$\ and $-l_{0}$.
The photon pair is initially prepared in an extended Werner-like state defined as
\begin{align}\label{S0}
\rho^{(0)}=\frac{1-\gamma}{4}I+\gamma\left\vert \Psi_{0}\right\rangle \left\langle\Psi_{0}\right\vert,
\end{align}
where $0\leq\gamma\leq1$ denotes the purity of the initial state and $\left\vert \Psi_{0}\right\rangle$ is the Bell-like state given by%
\begin{align}
\left\vert \Psi_{0}\right\rangle =\cos\left(\frac{\theta}{2}\right)\left\vert l_{0},-l_{0}\right\rangle
+e^{i\phi}\sin\left(\frac{\theta}{2}\right)\left\vert -l_{0},l_{0}\right\rangle,
\end{align}
with $0\leq\theta\leq\pi$ and $0\leq\phi\leq2\pi$.
It is worth noting that the quantum state (\ref{S0}) recovers the Bell state considered in Ref.~\cite{leonhard2015universal} for purity $\gamma=1$ and $\theta=\pi/2$.
The extended Werner-like states play an important role in many applications of QIS~\cite{horodecki1999general,acn2006from}.
The action of the turbulent atmosphere on the photons can be treated as a linear map $\Lambda$,
in terms of which the received state at the detectors reads as
\begin{align}\label{S}
\rho=\left(\Lambda_{1}\otimes\Lambda_{2}\right)\rho^{(0)},
\end{align}
where $\Lambda_{1}$ and $\Lambda_{2}$ are the actions of the atmospheric turbulence on the individual photon state.
The density matrix of the photonic state (\ref{S}) can be constructed as~\cite{bellomo2007nonmarkovian}
\begin{align}\label{S1}
\rho \propto& \sum_{ii^{\prime},jj^{\prime}}\rho_{\left\vert ij\right\rangle\left\langle i^{\prime}j^{\prime}\right\vert}\left\vert ij\right\rangle\left\langle i^{\prime}j^{\prime}\right\vert \notag \\
=& \sum_{ii^{\prime},jj^{\prime}}\rho_{ii^{\prime},jj^{\prime}}\left\vert ij\right\rangle \left\langle i^{\prime}j^{\prime}\right\vert \notag \\
=& \sum_{ii^{\prime},jj^{\prime}}\sum_{ll^{\prime},mm^{\prime}}\Lambda_{1ii^{\prime}}^{ll^{\prime}}\Lambda_{2jj^{\prime}}^{mm^{\prime}}\rho_{ll^{\prime},
mm^{\prime}}^{\left(0\right)}\left\vert ij\right\rangle\left\langle i^{\prime}j^{\prime}\right\vert \notag \\
=& \sum_{ii^{\prime},jj^{\prime}}\sum_{ll^{\prime},mm^{\prime}}\Lambda_{1ii^{\prime}}^{ll^{\prime}}\Lambda_{2jj^{\prime}}^{mm^{\prime}}\rho_{\left\vert lm\right\rangle \left\langle l^{\prime}m^{\prime}\right\vert }^{\left(0\right)}\left\vert ij\right\rangle \left\langle i^{\prime}j^{\prime}\right\vert.
\end{align}
Here, we let $\Lambda_{1}=\Lambda_{2}=\Lambda$ due to the same effect of turbulent atmosphere on the photons.
The elements $\Lambda_{l,l^{\prime}}^{l_{0},l_{0}^{\prime}}=\sum_{p}\Lambda_{pl,pl^{\prime}}^{p_0l_{0},p_0^{\prime}l_{0}^{\prime}}$ of the linear map $\Lambda$ is given by~\cite{leonhard2015universal}
\begin{align}\label{Lambda}
\Lambda_{l,\pm l^{\prime}}^{l_{0},l_{0}^{\prime}}=\frac{\delta_{l_{0}-l_{0}^{\prime},l\mp l}}{2\pi}
\int_{0}^{\infty}\mathrm{d}rR_{p_0l_{0}}\left(r\right)R_{p_0l_{0}}^{\ast}\left(r\right)r\times \nonumber\\
\int_{0}^{2\pi}\mathrm{d}\vartheta
\mathrm{e}^{-\mathrm{i}\vartheta \left[l\pm l-\left(l_{0}+l_{0}^{\prime}\right)\right]/2}
\mathrm{e}^{-D_{\phi}\left(2r\left\vert \sin\left(\vartheta/2\right)\right\vert \right)/2},
\end{align}
where
\begin{align}
R_{p_0l_{0}}\left(r\right)=&\frac{2}{\omega_{0}}\sqrt{\frac{p_0!}{(p_0+\left\vert l_{0}\right\vert)!}}
\left(\frac{r\sqrt{2}}{\omega_{0}}\right)^{\left\vert l_{0}\right\vert}
L_{p_0}^{\left\vert l_{0}\right\vert }\left(\frac{2r^{2}}{\omega_{0}}\right)\times\nonumber\\
&\exp\left(-\frac{r^{2}}{\omega_{0}^{2}}\right),
\end{align}
is the radial part of LG beam at propagation distance $z=0$ \cite{gopaul2007effect} with generalized Laguerre polynomials
\begin{align}
L_{p_0}^{\left\vert l_{0}\right\vert}\left(x\right)=\sum_{m=0}^{p_0}\left(-1\right)^{m}\frac{\left(\left\vert l_{0}\right\vert +p_0\right)!}
{\left(p_0-m\right)!\left(\left\vert l_{0}\right\vert +m\right)!m!}x^{m}.
\end{align}
Here, we consider $D_{\phi}=6.88\left(r/r_{0}\right)^{5/3}$, which is the phase structure function of the Kolmogorov model of turbulence with the Fried parameter
\begin{align}
r_{0}=\left(0.423C_{n}^{2}k^{2}L\right)^{-3/5},
\end{align}
where $C_{n}^{2}$ is the index-of-refraction structure constant, $L$ is the propagation distance,
and $k$ is the optical wave number~\cite{andrews2005}.
The density matrix of the input state (\ref{S0}) in the basis $\left\{\left\vert l_{0},l_{0}\right\rangle ,\left\vert l_{0},-l_{0}\right\rangle ,\left\vert
-l_{0},l_{0}\right\rangle ,\left\vert -l_{0},-l_{0}\right\rangle\right\}$ can be written as an X form
\begin{align}
\rho^{\left(0\right)}=\left(
\begin{array}{cccc}
\rho_{11}^{\left(0\right)} &  &  & \rho_{14}^{\left(0\right)} \\
& \rho_{22}^{\left(0\right)} & \rho_{23}^{\left(0\right)} &  \\
& \rho_{32}^{\left(0\right)} & \rho_{33}^{\left(0\right)} &  \\
\rho_{41}^{\left(0\right)} &  &  & \rho_{44}^{\left(0\right)}
\end{array}
\right),
\end{align}
where
\begin{align}
\rho_{11}^{\left(0\right)}=& \frac{1-\gamma}{4},\rho_{22}^{\left(0\right)}=\frac{1-\gamma}{4}+\gamma\cos^{2}\left(\frac{\theta}{2}\right),\notag \\
\rho_{33}^{\left(0\right)}=& \frac{1-\gamma}{4}+\gamma\sin^{2}\left(\frac{\theta}{2}\right),\rho_{44}^{\left(0\right)}=\frac{1-\gamma}{4},\notag \\
\rho_{14}^{\left(0\right)}=& 0,\rho_{23}^{\left(0\right)}=\frac{1-\gamma}{4}+\frac{\gamma}{2}e^{-i\phi}\sin\theta,\notag \\
\rho_{41}^{\left(0\right)}=& \rho_{14}^{\left(0\right)\ast},\rho_{32}^{\left(0\right)}=\rho_{23}^{\left(0\right)\ast}.
\end{align}
According to Eq.~(\ref{S1}), the normalized density matrix of the output state (\ref{S}) can also be expressed in the X form as
\begin{align}
\rho=&\frac{\sum_{ii^{\prime},jj^{\prime}}\rho_{\left\vert ij\right\rangle\left\langle i^{\prime}j^{\prime}\right\vert}\left\vert ij\right\rangle\left\langle i^{\prime}j^{\prime}\right\vert}{\mathrm{Tr}\left(\sum_{ii^{\prime},jj^{\prime}}\rho_{\left\vert ij\right\rangle\left\langle i^{\prime}j^{\prime}\right\vert}\left\vert ij\right\rangle\left\langle i^{\prime}j^{\prime}\right\vert\right)}\notag \\
=&\left(
\begin{array}{cccc}
\rho_{11} &  &  & \rho_{14} \\
& \rho_{22} & \rho_{23} &  \\
& \rho_{32} & \rho_{33} &  \\
\rho_{41} &  &  & \rho_{44}
\end{array}
\right),
\end{align}
with
\begin{align}\label{X}
\rho_{11}=& \left(a^{2}\rho_{11}^{\left(0\right)}+ab\rho_{22}^{\left(0\right)}+ab\rho_{33}^{\left(0\right)}+b^{2}\rho_{44}^{\left(0\right)}\right)/\left(a+b\right)^{2},  \notag\\
\rho_{22}=& \left(ab\rho_{11}^{\left(0\right)}+a^{2}\rho_{22}^{\left(0\right)}+b^{2}\rho_{33}^{\left(0\right)}+ab\rho_{44}^{\left(0\right)}\right)/\left(a+b\right)^{2},  \notag\\
\rho_{33}=& \left(ab\rho_{11}^{\left(0\right)}+b^{2}\rho_{22}^{\left(0\right)}+a^{2}\rho_{33}^{\left(0\right)}+ab\rho_{44}^{\left(0\right)}\right)/\left(a+b\right)^{2},  \notag\\
\rho_{44}=& \left(b^{2}\rho_{11}^{\left(0\right)}+ab\rho_{22}^{\left(0\right)}+ab\rho_{33}^{\left(0\right)}+a^{2}\rho_{44}^{\left(0\right)}\right)/\left(a+b\right)^{2},  \notag\\
\rho_{14}=& a^{2}\rho_{14}^{\left(0\right)}/\left(a+b\right)^{2},\rho_{41}=a^{2}\rho_{41}^{\left(0\right)}/\left(a+b\right)^{2}, \notag\\
\rho_{23}=& a^{2}\rho_{23}^{\left(0\right)}/\left(a+b\right)^{2},\rho_{32}=a^{2}\rho_{32}^{\left(0\right)}/\left(a+b\right)^{2},
\end{align}
where
\begin{align}\label{ab}
a =& \Lambda_{l_{0},l_{0}}^{l_{0},l_{0}}=\Lambda_{-l_{0},-l_{0}}^{-l_{0},-l_{0}}=\Lambda_{-l_{0},l_{0}}^{-l_{0},l_{0}}=\Lambda_{l_{0},-l_{0}}^{l_{0},-l_{0}},
\notag \\
b =& \Lambda_{l_{0},l_{0}}^{-l_{0},-l_{0}}=\Lambda_{-l_{0},-l_{0}}^{l_{0},l_{0}}.
\end{align}
In this sense, the output OAM state can be treated as two-qubit.
Then, we can express the entanglement by Wootters' concurrence~\cite{wootters1998entanglement} for the X state as
\begin{align}
C\left(\rho\right)=\max\left\{ 0,\left\vert\rho_{14}\right\vert-\sqrt{\rho_{22}\rho_{33}},\left\vert\rho_{23}\right\vert-\sqrt{\rho_{11}\rho_{44}}\right\}.
\end{align}
With Eqs. (\ref{X}) and (\ref{ab}), an analytical form of the concurrence can be derived as
\begin{align}
C\left(\rho\right)=\max\left\{0,\frac{a^{2}\gamma\sin\theta-2ab\gamma}{\left( a+b\right)^{2}}-\frac{1-\gamma}{2}\right\}.
\end{align}
For convenience, one can introduce the phase correlation length $\xi\left(l_{0}\right)$ which is defined as the average distance between the points in
the LG beam cross-section that have a phase difference of $\pi/2$. The phase correlation length can be expressed as~\cite{leonhard2015universal}
\begin{align}\label{cl}
\xi\left(l_{0}\right) =\sin\left(\frac{\pi}{2\left\vert l_{0}\right\vert}\right)
\frac{\omega_{0}}{2}\frac{\Gamma\left(\left\vert l_{0}\right\vert+3/2\right)}{\Gamma\left(\left\vert l_{0}\right\vert+1\right)},
\end{align}
with $\Gamma\left(x\right)$ the Gamma function .

\begin{figure}[tbh!]
\centering
\includegraphics[angle=0,width=4cm]{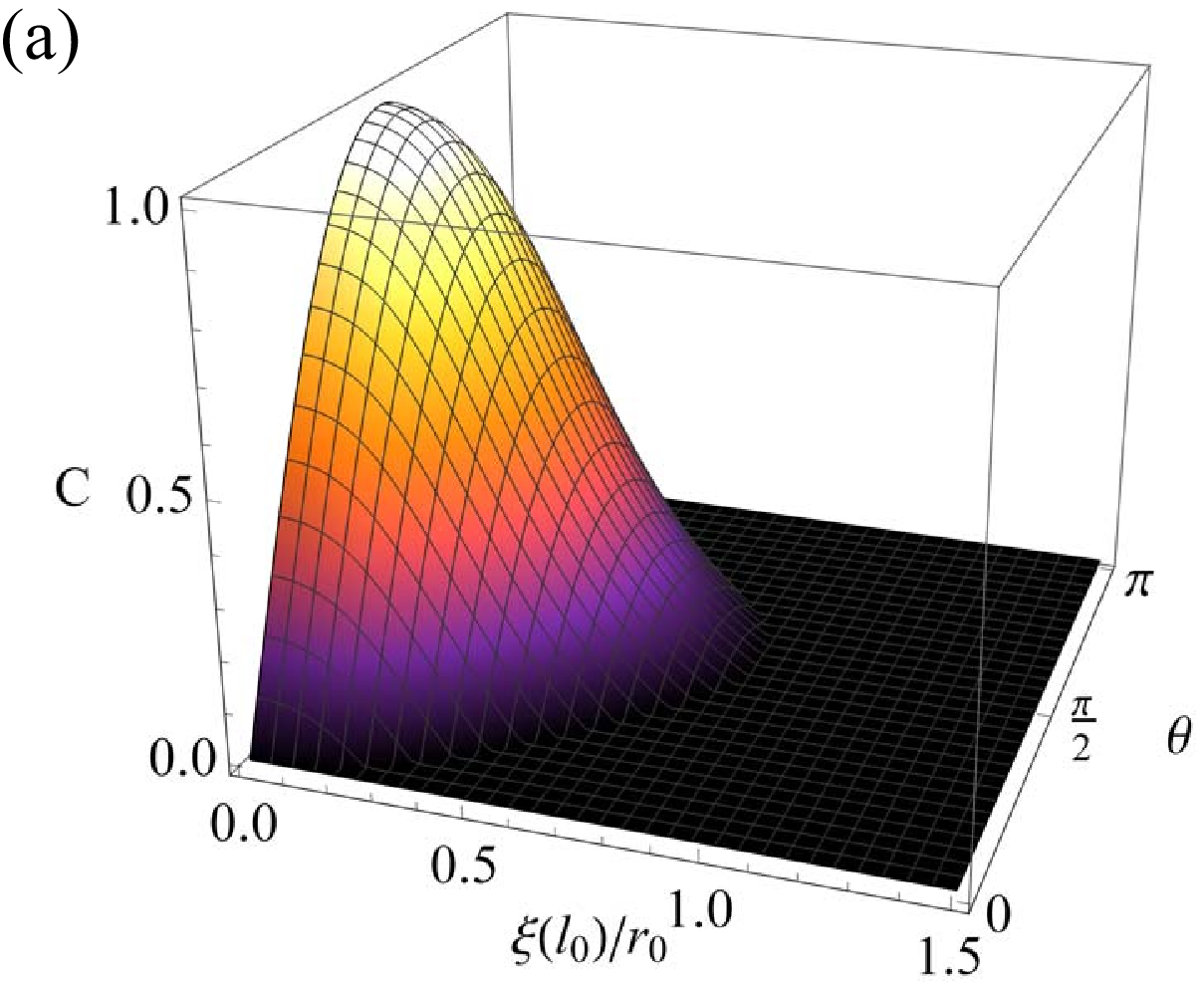}
\includegraphics[angle=0,width=4cm]{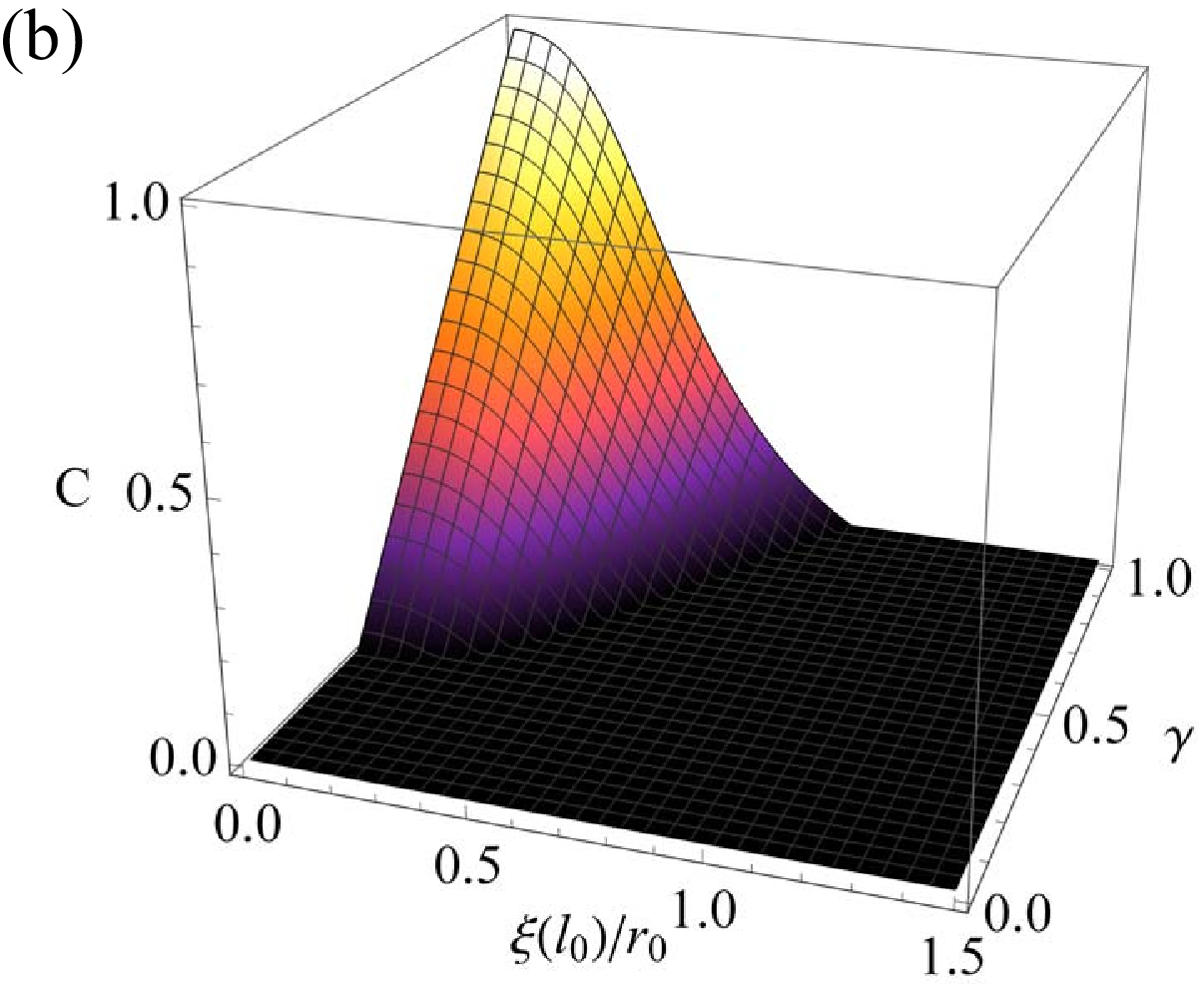}
\caption{Concurrence as a function of (a) $\xi(l_{0})/r_{0}$ and $\theta$, and of (b) $\xi(l_{0})/r_{0}$ and $\gamma$.
The parameters are chosen as $l_{0}=1$, $\omega_{0}=1$ for (a) $\gamma=1$ and (b) $\theta=\pi/2$.}
\label{fig:fig2}
\end{figure}

Then, we plot the concurrence $C\left(\rho\right)$ as a function of the ratio $x=\xi\left(l_{0}\right)/r_{0}$ and the initial state parameter $\theta$ in Fig.~\ref{fig:fig2}(a).
We can see that the concurrence decays with the increase of the ratio $\xi\left(l_{0}\right)/r_{0}$ and decreases fast to zero with non-asymptotical vanishing,
the phenomenon of which is termed as entanglement sudden death (ESD)~\cite{yu2006quantum,yu2009sudden}.
Moreover, we plot the concurrence as a function of the ratio $\xi\left(l_{0}\right)/r_{0}$ and the purity parameter $\gamma$ in Fig.~\ref{fig:fig2}(b).
It is found that the concurrence is vanishing when the initial state is unentangled for $0\leq\gamma\leq1/\left(1+2\sin\theta\right)$,
since no entanglement can be created between the two photons under independent atmospheric turbulences for initially disentangled states.
In the next part, we will investigate the properties of quantum coherence and quantum correlation via relative entropy of coherence and LQU,
which may exhibit qualitative differences from that of entanglement via concurrence.

\section{Quantum coherence and quantum correlation of OAM state in atmospheric turbulence}\label{sec:sec3}

We can quantify the degree of quantum coherence and quantum correlation via relative entropy of coherence~\cite{baumgratz2014quantifying} and LQU \cite{girolami2013characterizing}.
The relative entropy of coherence for a quantum state $\rho$ is given by~\cite{baumgratz2014quantifying}
\begin{align}\label{C}
\mathcal{C}_{\mathrm{rel.ent.}}\left(\rho\right)=S\left(\rho_{\mathrm{diag}}\right)-S\left(\rho\right),
\end{align}
where $S\left(\rho\right)=-\mathrm{Tr\left(\rho\log_{2}\rho\right)}$ is the von Neumann entropy and $\rho_{\mathrm{diag}}$ is the ``closest'' non-coherent state, which is diagonal by deleting all the off-diagonal elements of $\rho$. The relative entropy of coherence is shown to be nice coherence quantifier, fulfilling all the conditions proposed in Ref.~\cite{Streltsov2017} and exactly computable for bipartite quantum state.

The LQU, a measure of discord-like quantum correlation, is the minimal quantum uncertainty achievable on a single local measurement~\cite{girolami2013characterizing}.
For a bipartite system $AB$ with the state $\rho$ subject to a local measurement on the subsystem $A$, quantum uncertainty will yield if the state $\rho$ is nonclassical.
Then the LQU is defined by the minimal Wigner-Yanase skew information~\cite{wigner1963information} as
\begin{align}\label{LQU}
\mathrm{LQU}(\rho)=&\min_{\{K_{A}\}}\{\mathcal{I}\left(\rho,K_{A}\otimes I_{B}\right)\}\notag\\
=&-\frac{1}{2}\min_{\{K_{A}\}}\{\mathrm{Tr}\left([\rho,K_{A}\otimes I_{B}]^{2}\right)\},
\end{align}
where $K_{A}$ is a local observable on system $A$ and $I_{B}$ is the identity operator of subsystem $B$.
The LQU has a simple expression for a $2\times d$ quantum system as
\begin{align}
\mathrm{LQU}(\rho)=1-\lambda_{\max}\left\{W_{AB}\right\},
\end{align}
where $\lambda_{\max}$ is the maximal eigenvalue of the $3\times3$ matrix $W_{AB}$ with the elements
\begin{align}
\left(W_{AB}\right)_{ij}=\mathrm{Tr}\left[\sqrt{\rho_{AB}}\left(\sigma_{iA}\otimes I_{B}\right)\sqrt{\rho_{AB}}\left(\sigma_{jA}\otimes I_{B}\right)\right],
\end{align}
with $\sigma_{iA}\left(i=x,y,z\right)$ denoting the Pauli matrixes of subsystem $A$. LQU satisfies all the known bona fide criteria for a discord-like quantifier and is a sufficient
resource for phase estimation in quantum metrology~~\cite{girolami2013characterizing}.

\begin{figure}[tbh!]
\centering
\includegraphics[angle=0,width=4cm]{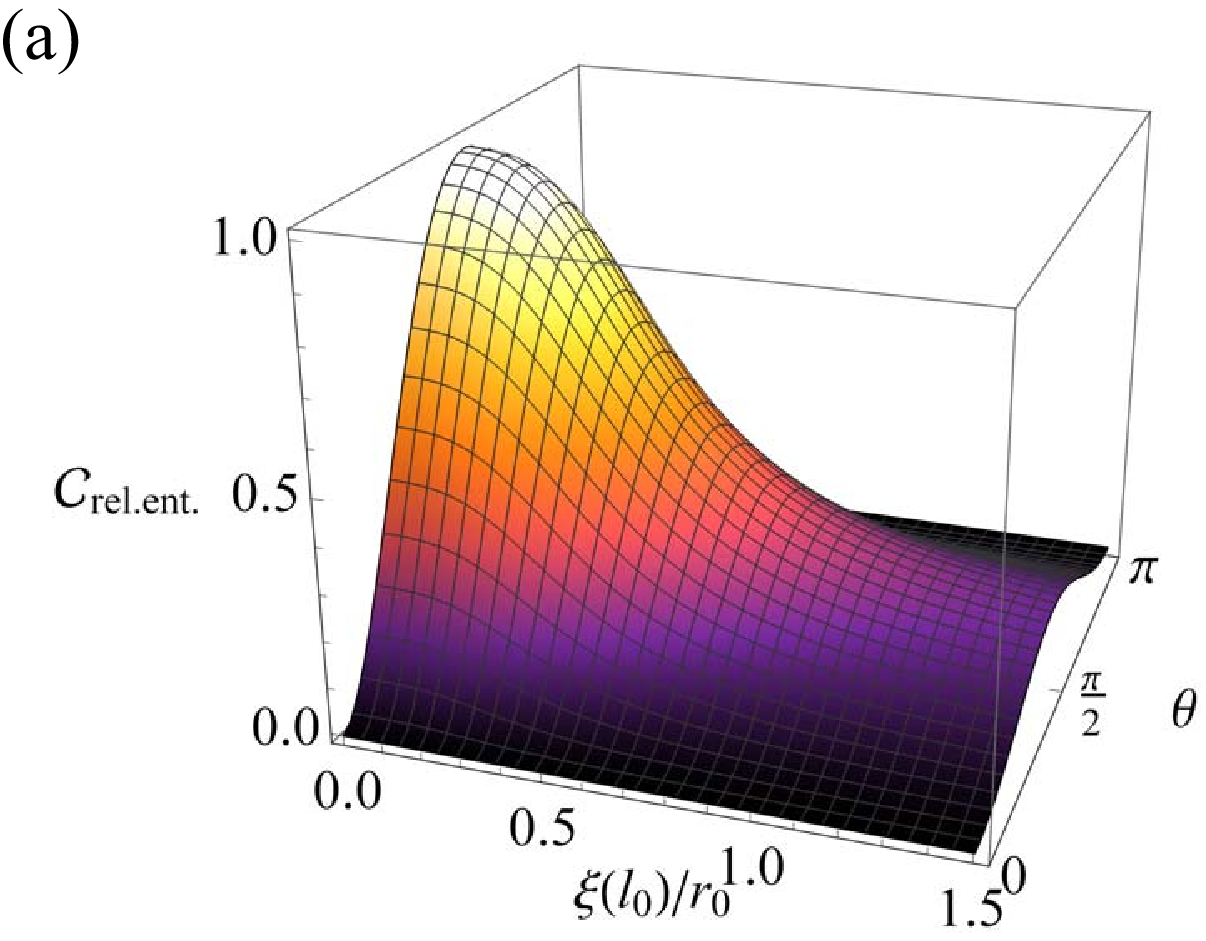}
\includegraphics[angle=0,width=4cm]{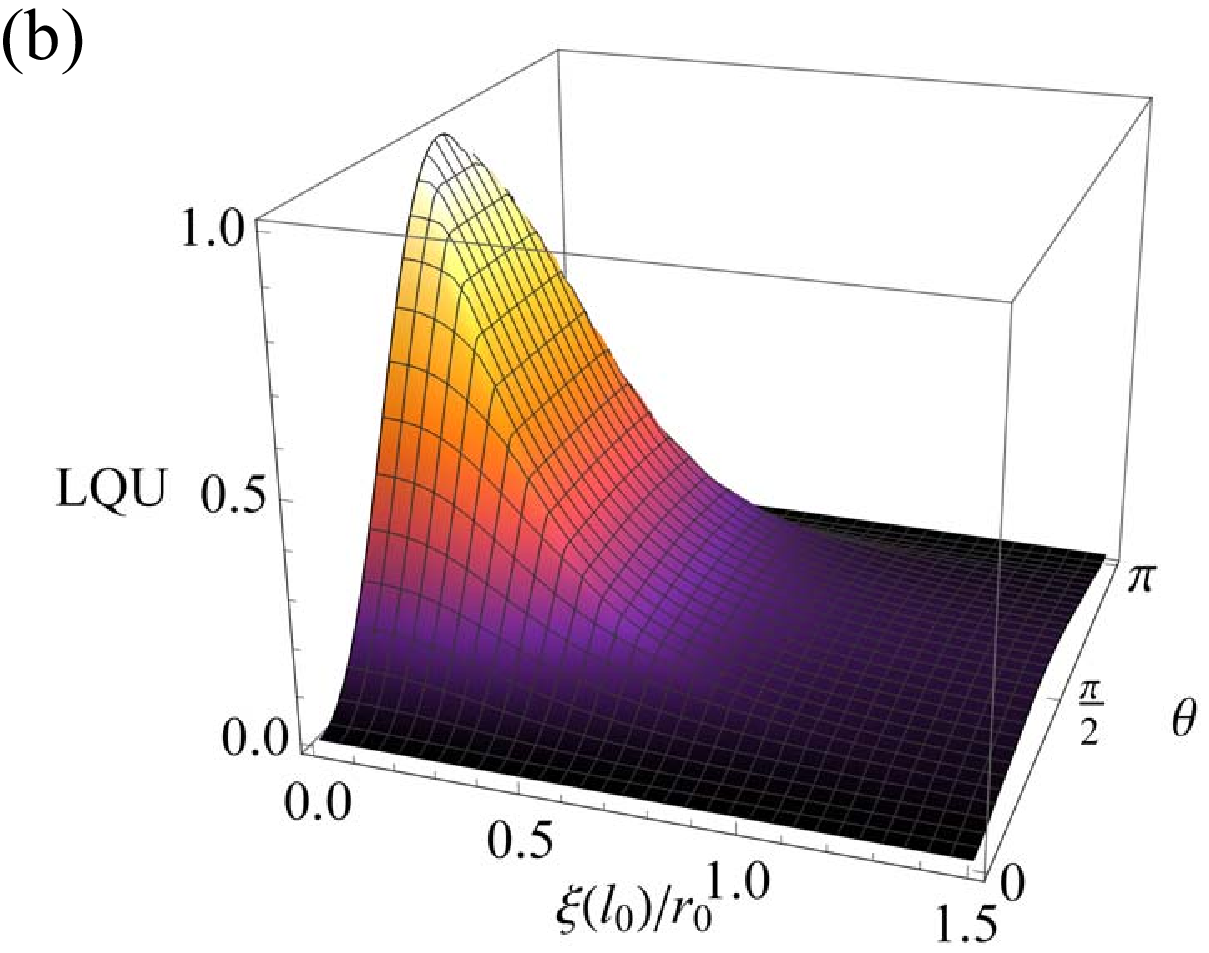}
\includegraphics[angle=0,width=4cm]{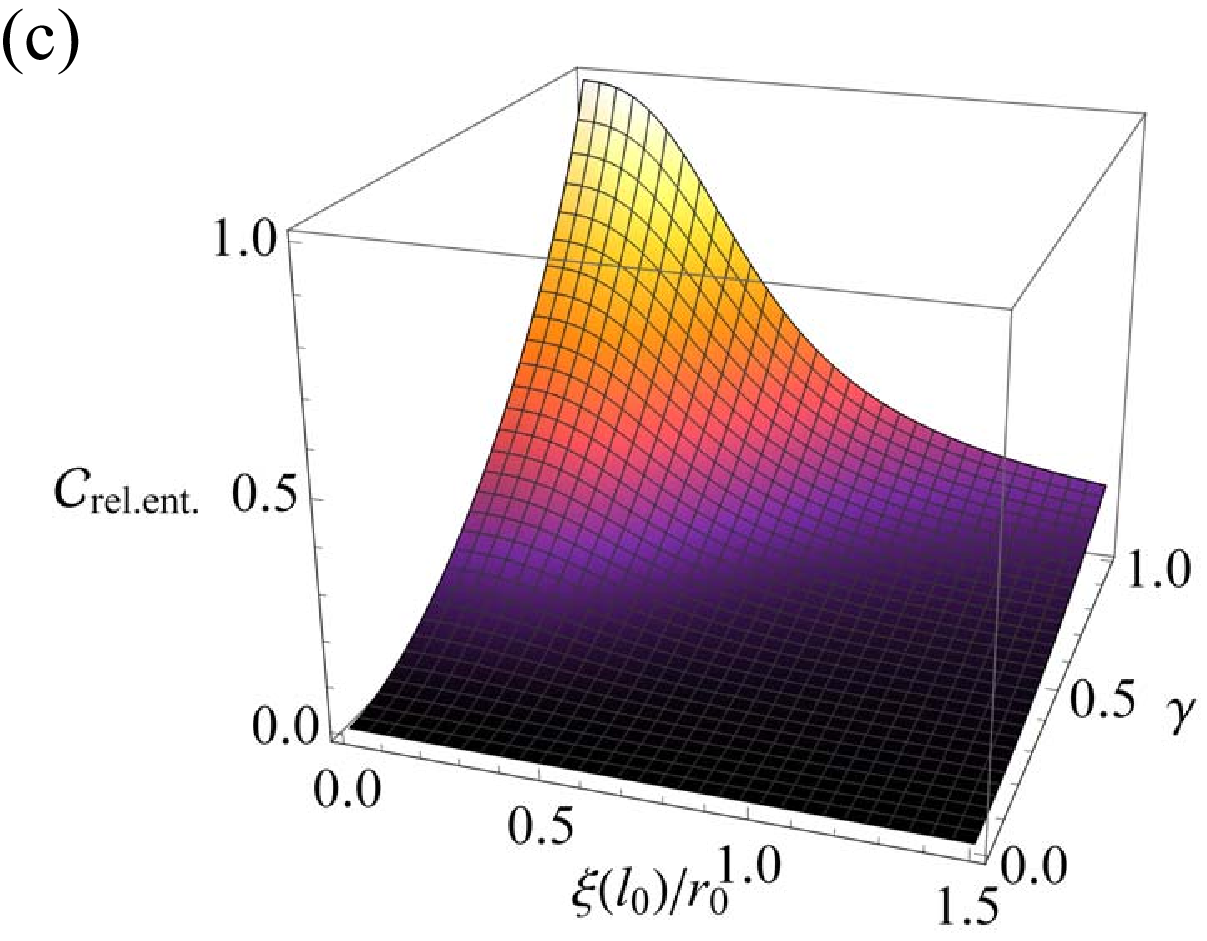}
\includegraphics[angle=0,width=4cm]{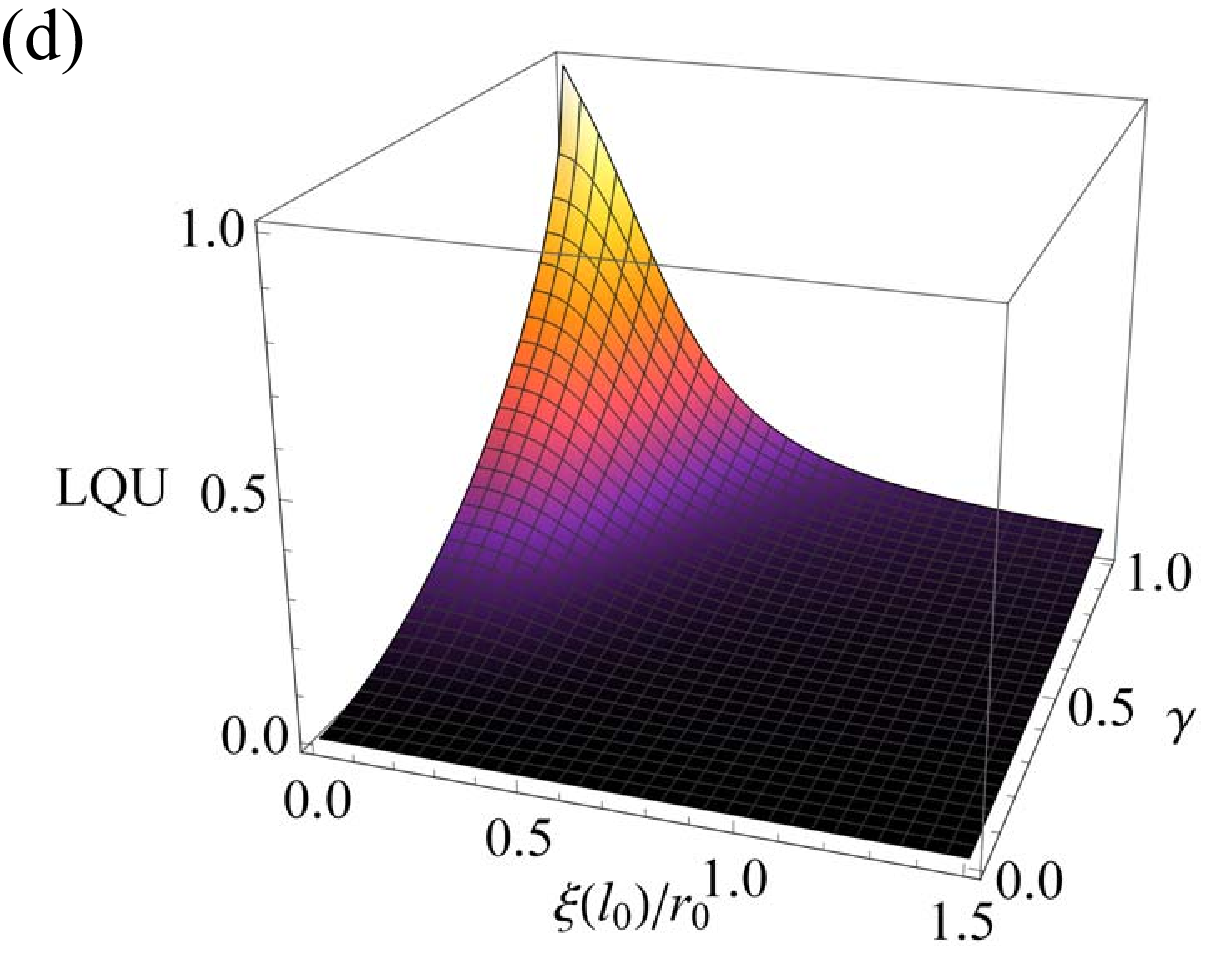}
\caption{(a, c) Relative entropy of coherence and (b, d) LQU as functions of (a, b) $\xi(l_{0})/r_{0}$ and $\theta$, and of (c, d) $\xi(l_{0})/r_{0}$ and $\gamma$.
The parameters are chosen as $l_{0}=1$, $\omega_{0}=1$ for (a, b) $\gamma=1$ and (c, d) $\theta=\pi/2$.}
\label{fig:fig3}
\end{figure}
Then, the relative entropy of coherence and LQU as functions of $\xi\left(l_{0}\right)/r_{0}$ and $\theta$ are displayed in Fig.~\ref{fig:fig3}(a) and Fig.~\ref{fig:fig3}(b).
We can see that the relative entropy of coherence and LQU decay fast with the increase of the ratio $\xi\left(l_{0}\right)/r_{0}$ when $\theta$ is close to $\pi/2$ and then decrease slowly in a non-vanishing manner even when the $\xi\left(l_{0}\right)/r_{0}$ is large enough.
Therefore, the phenomenon of sudden vanishing as for the entanglement (ESD) does not occur here.
We also show the relative entropy of coherence and LQU as functions of $\xi\left(l_{0}\right)/r_{0}$ and $\gamma$ in Fig.~\ref{fig:fig3}(c) and Fig.~\ref{fig:fig3}(d).
When the purity $\gamma$ of the initial state is close to zero,
the quantum coherence and quantum correlation of the two qubits are very weak but still be non-vanishing,
which is quite different from the entanglement shown in Fig.~\ref{fig:fig2}(b).
When $\gamma$ is close to $1$, the relative entropy of coherence and LQU seem to decay in a nearly exponential manner with the increase of $\xi\left(l_{0}\right)/r_{0}$,
which is also contrast to that of entanglement with sudden vanishing as shown in Fig.~\ref{fig:fig2}(b).

\begin{figure}[tbh!]
\centering
\includegraphics[angle=0,width=4cm]{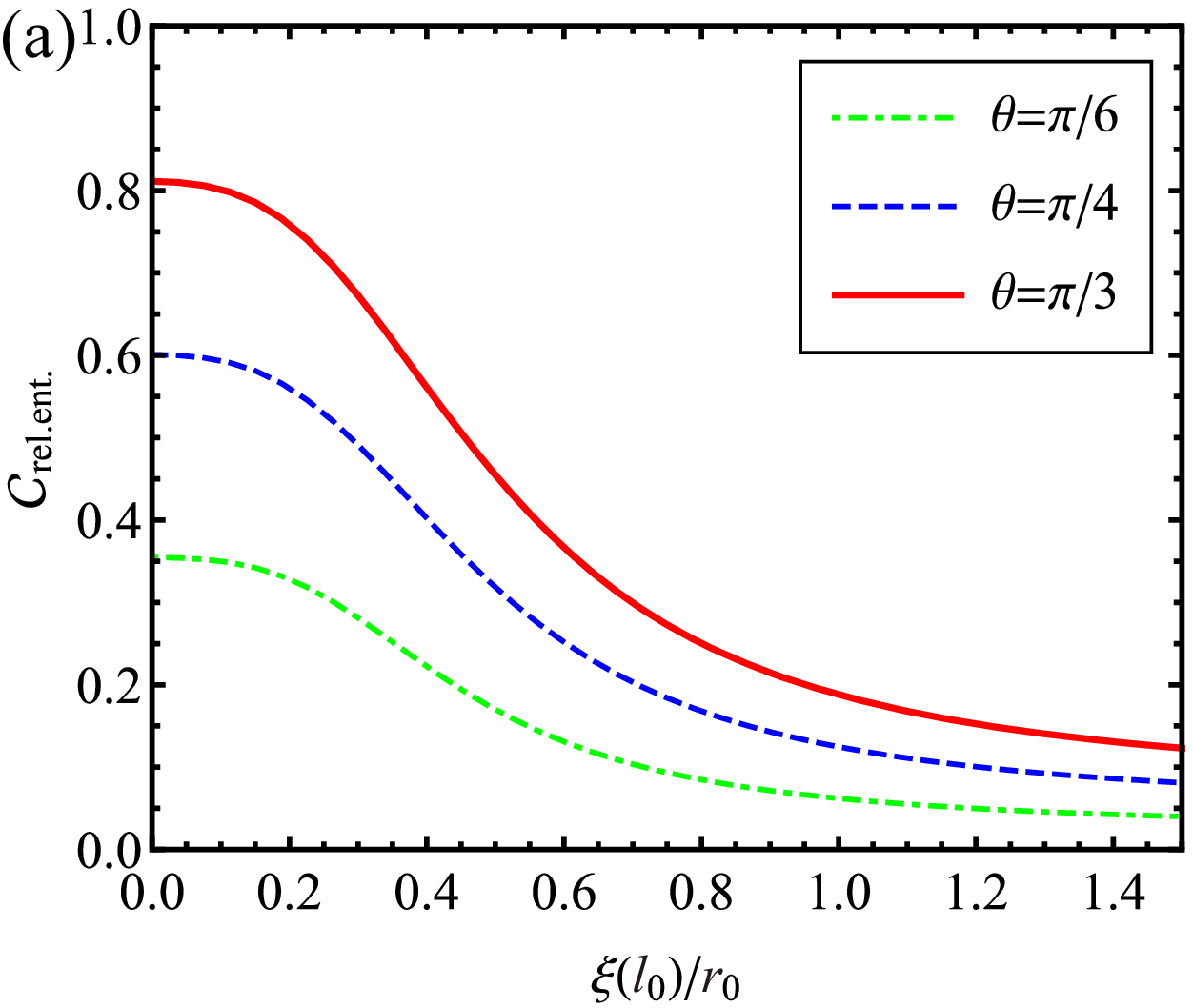}
\includegraphics[angle=0,width=4cm]{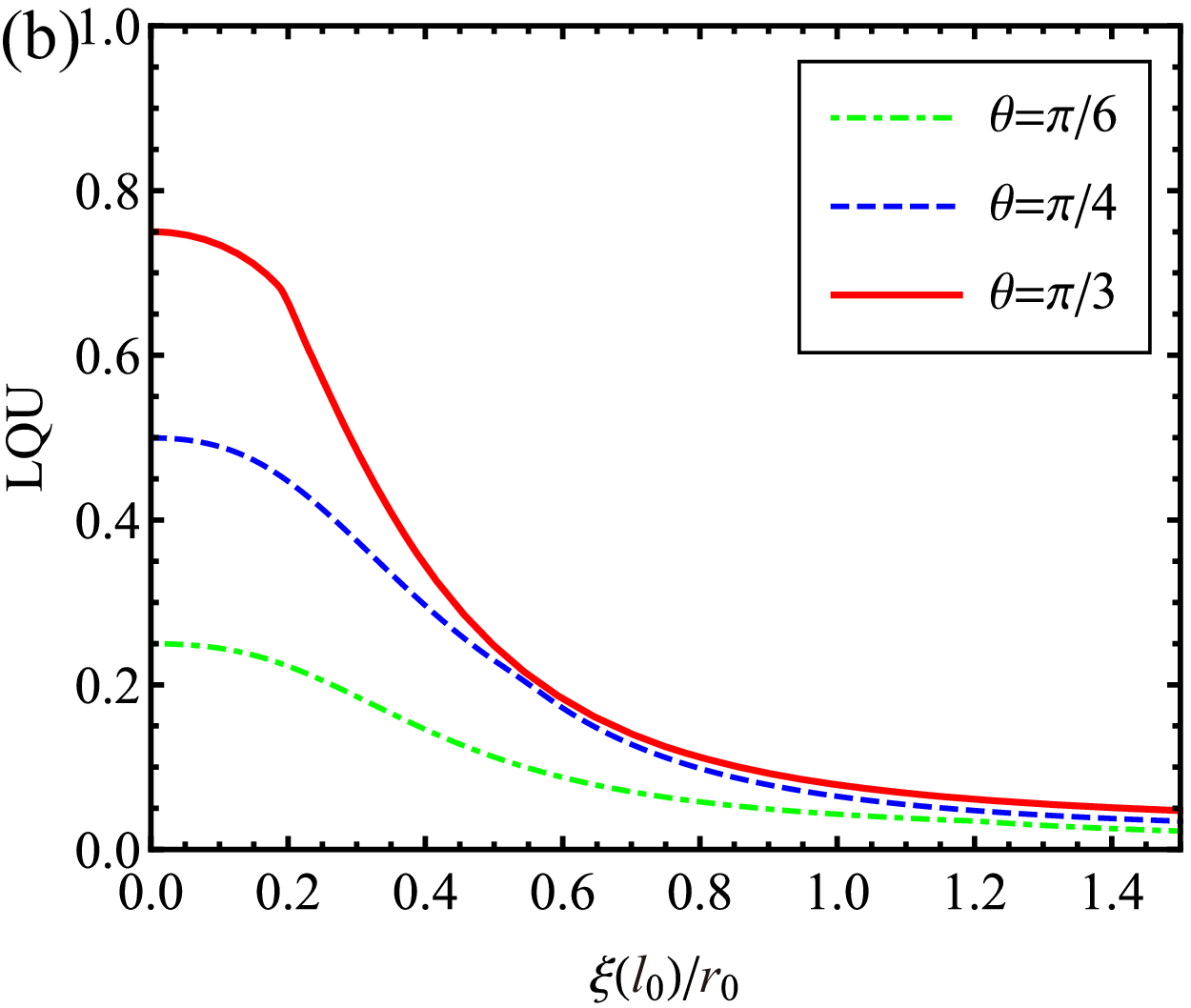}
\caption{(a) Relative entropy of coherence and (b) LQU as functions of $\xi(l_{0})/r_{0}$.
The parameters are chosen as $l_{0}=1$, $\omega_{0}=1$ and $\gamma=1$.}
\label{fig:fig4}
\end{figure}
In order to see the evolution in turbulent atmosphere more clearly,
the relative entropy of coherence and LQU as functions of $\xi\left(l_{0}\right)/r_{0}$ for different $\theta$ are also demonstrated in Fig.~\ref{fig:fig4}.
We can clearly see that the relative entropy of coherence and LQU decay more and more slowly with the increase of the ratio $\xi\left(l_{0}\right)/r_{0}$.
For certain initial state, for instance $\theta=\pi/3$\ shown in Fig.~\ref{fig:fig4}(b),
the decay rate of LQU may be non-continuously changed when the $\xi\left(l_{0}\right)/r_{0}$ is still small,
which is termed as the sudden change phenomenon~\cite{maziero2009classical,xu2010experimental} as for discord-like quantum correlations.

\begin{figure}[tbh!]
\centering
\includegraphics[angle=0,width=4cm]{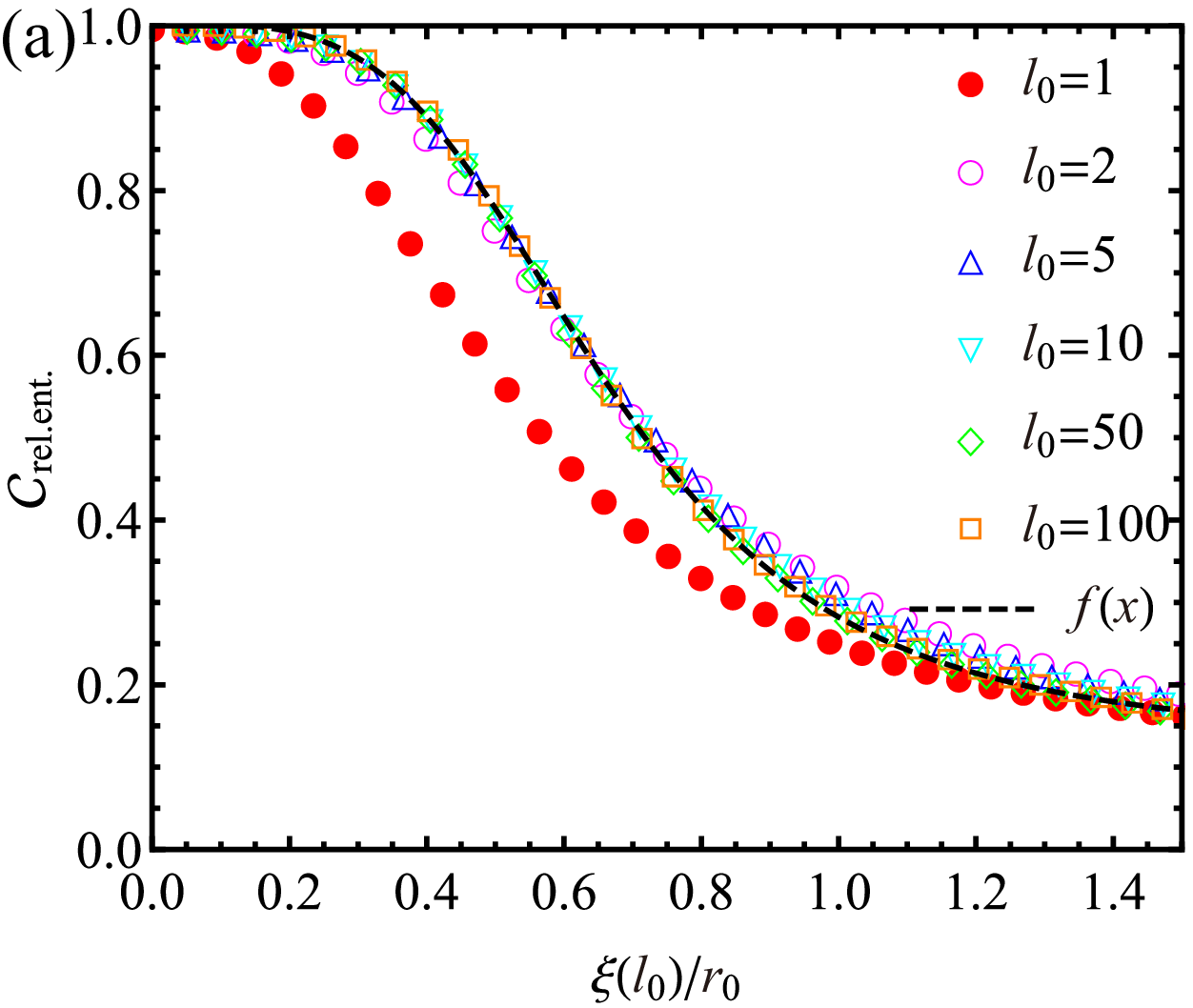}
\includegraphics[angle=0,width=4cm]{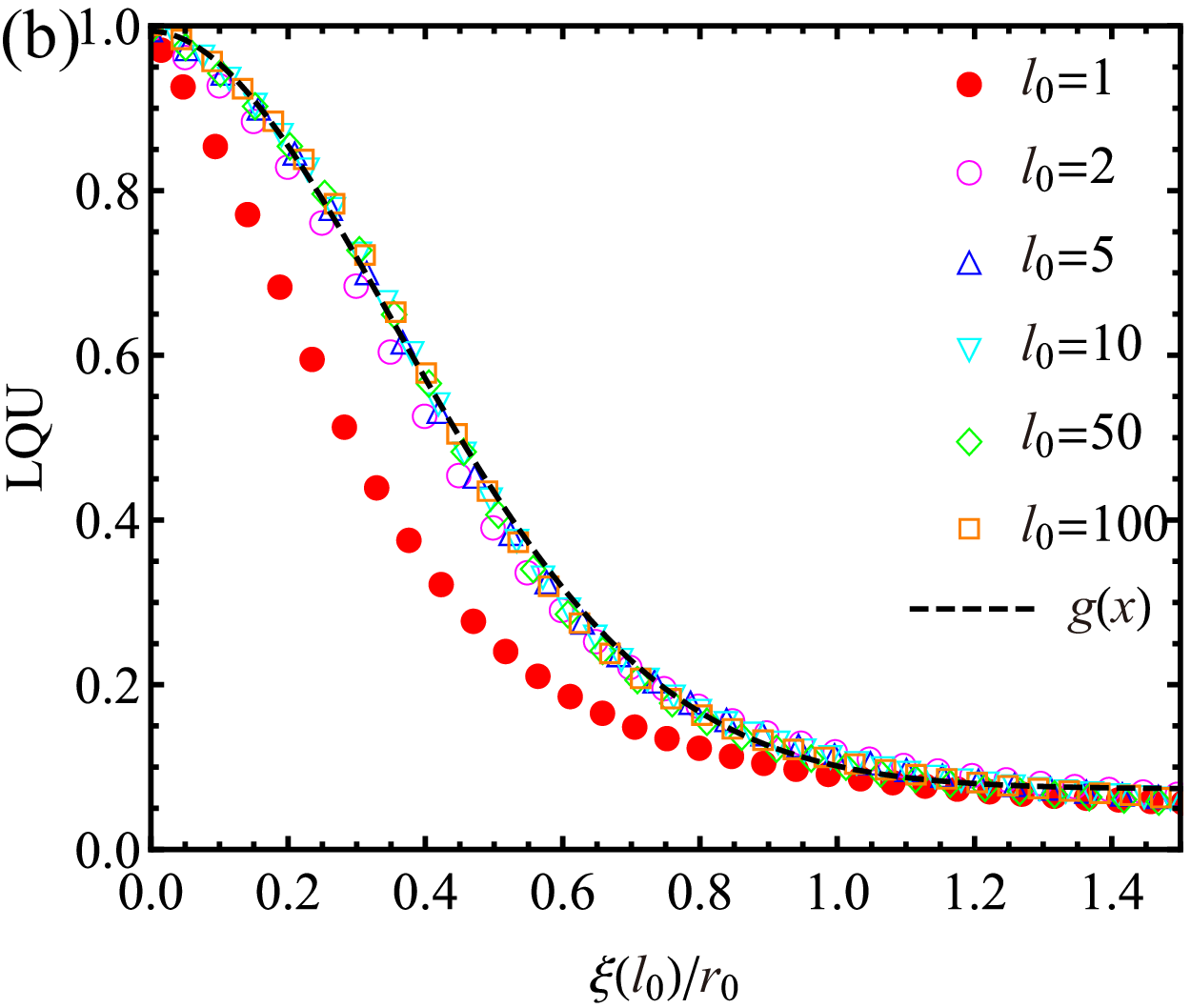}
\caption{(a) Relative entropy of coherence and (b) LQU as functions of $\xi(l_{0})/r_{0}$ under different $l_{0}$.
The parameters are chosen as $\omega _{0}=1$, $\theta=\pi/2$ and $\gamma=1$.}
\label{fig:fig5}
\end{figure}
Moreover, we would like to explore the precise decay laws of quantum coherence and quantum correlation for different values of phase correlation length which is determined by the azimuthal quantum number $l_{0}$ as shown in Eq.~(\ref{cl}).
As can be seen from Fig.~\ref{fig:fig5}(a) and Fig.~\ref{fig:fig5}(b), both the relative entropy and LQU decay fastest at $l_{0}=1$.
When the azimuthal quantum number $l_0$ increases, the decays of relative entropy of coherence and LQU are slowed down and finally collapse onto two universal curves. Their functions of the ratio $x=\xi\left(l_{0}\right)/r_{0}$ is given by $f(x)=\frac{0.183}{x^{3.78}+0.21}+0.131$ and $g(x)=0.92[\mathrm{exp}(-3.50x^{1.90})+0.08]$, which means that the relative entropy of coherence decays in a polynomial manner while the LQU decays in an exact exponential manner.
Since the LQU is a kind of nonclassical correlation apart from entanglement,
then it is natural to derive the universal exponential decay similar to that of entanglement reported in Ref.~\cite{leonhard2015universal}.
It is believed that nonclassical correlations (such as quantum discord and entanglement) reveal only parts of quantumness, while quantum coherence is a more general concept for quantumness~\cite{Streltsov2017}, which may be the physical mechanism underlined the phenomenon of the more robust decay law for coherence against atmospheric turbulence.

\section{Conclusions}\label{sec:sec4}

In conclusion, we have investigated the decay properties of quantum coherence and nonclassical correlations (entanglement and discord) for photonic states carrying orbital angular momentum (OAM)
through Kolmogorov turbulent atmosphere via relative entropy of coherence, local quantum uncertainty (LQU), and concurrence, respectively.
By considering that the photonic OAM qubits, generated from a source, are initially prepared in an extended Werner-like state (partially entangled),
the decay effects of the turbulent atmosphere are explored for the output state received by the detectors.
It is shown that the quantumness measured by concurrence, relative entropy of coherence and LQU decays as the increase of the ratio of phase correlation length and the Fried parameter but with different phenomena for different measures.
The concurrence decays suddenly to zero with the so-called entanglement sudden death (ESD), while both the relative entropy of coherence and the LQU decay asymptotically.
For certain initial state, the LQU may demonstrate an extra sudden change phenomenon when the ratio of phase correlation length to Fried parameter is not large.
Moreover, we study the decays of quantum coherence and quantum correlation with different values of phase correlation length (the azimuthal quantum number), and find that two different universal decay laws emerge as the azimuthal quantum number becomes large.
The decay of LQU is universally in an exact exponential manner similar to that of entanglement already reported in Ref.~\cite{leonhard2015universal} but with asymptotic vanishing.
By contrast, the decay of relative entropy is merely polynomial, which illustrates that the quantum coherence can be more robust against atmospheric turbulence.

\section*{Acknowledgments}
This work is supported by the National Natural Science Foundation of China (NSFC) (Grant Nos.~11504140 and~11504139), the Natural Science Foundation of Jiangsu Province (Grant Nos.~BK20140128 and~BK20140167) and the Fundamental Research Funds for the Central Universities (Grant No.~JUSRP51517).

\end{document}